\documentclass[iop]{emulateapj-rtx4}

\usepackage{mathrsfs }
\usepackage{natbib}
\usepackage{amssymb}
\usepackage{ulem} 
\begin{document}

\title{Substructure within the SSA22 protocluster at $z\approx3.09$\altaffilmark{1}}

\author{
 Michael W. Topping,\altaffilmark{2}	
 Alice E. Shapley,\altaffilmark{2} \&
 Charles C. Steidel,\altaffilmark{3}
 }

\altaffiltext{1}{Based on data obtained at the W.M. Keck Observatory, which is operated as a scientific partnership among the California Institute of Technology, the University of California,  and the National Aeronautics and Space Administration, and was made possible by the generous financial support of the W.M. Keck Foundation.}
\altaffiltext{2}{Department of Physics and Astronomy, University of California, Los Angeles, 430 Portola Plaza, Los Angeles, CA 90095, USA}
\altaffiltext{3}{Cahill Center for Astronomy and Astrophysics, California Institute of Technology, 1216 East California Boulevard., MS 249-17, Pasadena, CA 91125, USA}

\email{mtopping@astro.ucla.edu}

\shortauthors{Topping et al.}


\shorttitle{Substructure within the SSA22 protocluster}


\begin{abstract} 
We present the results of a densely sampled spectroscopic survey of the SSA22 protocluster at $z\approx 3.09$.
Our sample with Keck/LRIS spectroscopy
includes 106 Ly$\alpha$ Emitters (LAEs) and 40 Lyman Break Galaxies (LBGs) at $z=3.05-3.12$.
These galaxies are contained within the $9'\times9'$ region in which the protocluster
was discovered, which also hosts the maximum galaxy overdensity
in the SSA22 region. The redshift histogram of our spectroscopic sample reveals
two distinct peaks, at $z=3.069$ (blue; 43 galaxies) and $z=3.095$ (red; 103 galaxies).
Furthermore, objects in the blue
and red peaks are segregated on the sky, with galaxies in the blue peak concentrating
towards the western half of the field. These results suggest that the blue and red
redshift peaks represent two distinct structures in physical space. Although the double-peaked
redshift histogram is traced in the same manner by LBGs and LAEs, and brighter and fainter galaxies,
we find that nine out of 10 X-ray AGNs in SSA22, and all seven spectroscopically-confirmed giant Ly$\alpha$ ``blobs,"
reside in the red peak. We combine our dataset
with sparsely sampled spectroscopy from the literature over a significantly wider area, finding
preliminary evidence that the
double-peaked structure in redshift space extends beyond the region of our dense spectroscopic
sampling. In order to fully characterize the three-dimensional structure, dynamics, and evolution
of large-scale structure in the SSA22 overdensity, we require the measurement of large samples of LAE and LBG
redshifts over a significantly wider area, as well as detailed comparisons with cosmological
simulations of massive cluster formation.
\end{abstract} 


\keywords{galaxies: clusters: individual (SSA22) --- galaxies: formation ---  galaxies: high-redshift ---
galaxies: starburst --- large-scale structure of universe}

\section{Introduction}
\label{sec:introduction}
Clusters represent the largest gravitationally bound
and densest structures in the universe. As such,
they provide a unique context for studying the formation
of both galaxies and large-scale structure. In order to understand
how galaxies form and evolve in dense environments, and
how these environments themselves evolve, it is important
to trace the origins of galaxy clusters. Protoclusters
are identified as regions containing significant overdensities
of galaxies, which may not yet be virialized but
will later evolve into rich galaxy clusters.
These cluster seeds provide insight into the origins of the environmental
trends observed among galaxies \citep[e.g.,][]{Dressler1980}, as well as the assembly of
clusters themselves. At this point, tens of protoclusters
have been spectroscopically confirmed at $z> 2$ \citep{Chiang2013},
with an order of magnitude more high-redshift protocluster candidates
photometrically identified \citep[e.g.,][]{Planck2015a,Chiang2014}.

The protocluster at $z\approx3.09$ in the SSA22 field is one
of the best-studied cluster progenitors in the literature.
This structure was first identified by \citet{Steidel1998}
in the course of a large spectroscopic survey for $z\sim 3$ Lyman
Break Galaxies \citep[LBGs;][]{Steidel2003}, and contains an LBG redshift-space
overdensity of $\delta_{LBG}=5.0\pm 1.2$ \citep{Steidel2000}.
Based on analytic calculations, \citet{Steidel1998} argued
that the SSA22 protocluster will evolve by $z=0$ into a rich Coma-like
cluster with $M\sim 10^{15} M_{\odot}$.

Following the initial protocluster discovery, an extensive suite of multi-wavelength observations has
been assembled in the SSA22 field, spanning from radio through X-ray
wavelengths and including high-resolution {\it Hubble Space Telescope}
imaging.  Specifically, mid- and far-infrared, submillimeter, and
millimeter observations have been used to study obscured
star formation and AGN activity in the SSA22 protocluster
\citep[e.g.,][]{Geach2005,Webb2009,Tamura2009,Umehata2015}, 
while deep JHK imaging has enabled the analysis of the most
massive protocluster galaxies \citep{Uchimoto2012,Kubo2013},
and deep {\it Chandra} observations have unveiled the X-ray AGN population
\citep{Lehmer2009a,Lehmer2009b}. The above observations suggest enhancements
in the star-formation rate density,
and the abundance of both massive galaxies and AGNs in the SSA22 protocluster, relative to
the ``field" at $z\sim 3$.

Quite remarkably, $\sim 1500$ narrowband-selected Ly$\alpha$
emitting galaxies (LAEs) at $z\sim 3.09$ have been identified in the original
$9'\times9'$ SSA22 pointing and the surrounding 1.38 deg$^2$ area
\citep{Steidel2000,Matsuda2004,Hayashino2004,Yamada2012a},
mapping the SSA22 protocluster and its surroundings on scales $>100$ comoving Mpc. The sample of SSA22 LAEs includes 12 giant
Ly$\alpha$ ``blobs" (LABs), i.e., regions of Ly$\alpha$ emission
extending over $100-200$ proper kpc in diameter 
\citep{Matsuda2011}, which may be specifically associated with galaxy overdensities
\citep{Prescott2008,Steidel2000}. Follow-up spectroscopy of LAEs in SSA22 has been used to
trace out the three-dimensional structure of this forming protocluster, in which
three 30 Mpc $\times$ 10 Mpc filaments of LAEs may intersect
in the region of highest galaxy overdensity \citep{Matsuda2005}.
However, with a sample of 56 spectroscopically-confirmed LAEs spread out over
170 arcmin$^2$, the sampling in redshift and position is not particularly dense.

We have obtained spatially dense spectroscopic sampling of LBGs
and LAEs, including 146 galaxies at  $z\sim 3.09$ within the original $9'\times9'$ pointing of \citet{Steidel1998}.
This region also contains the highest overdensity of LAEs in the
SSA22 structure \citep{Yamada2012a}. Our large spectroscopic sample reveals
previously unknown substructure within the SSA22 overdensity, both in redshift
space and on the sky. Characterizing the substructure in SSA22 is crucial for identifying
analogous protoclusters in cosmological simulations and understanding the origin
and fate of the SSA22 overdensity.

In \S\ref{sec:obsAndMethods}, we describe our observations and
redshift measurements for SSA22 LBGs and LAEs.
In \S\ref{sec:results} we discuss the substructure observed within the SSA22
protocluster both in redshift space and on the sky.
Finally, in \S\ref{sec:discussion} we compare with other work and discuss
the implications of our results for understanding the SSA22 protocluster.
For this work, we adopt cosmological parameters of
$H_0=67.8 \rm \ km \ s^{-1} \ Mpc^{-1}$, $\Omega_M = 0.3089$, and
$\Omega_{\Lambda}=0.6911$ \citep{Planck2015b}.

\begin{figure*}
\centering
\includegraphics[width=0.95\textwidth]{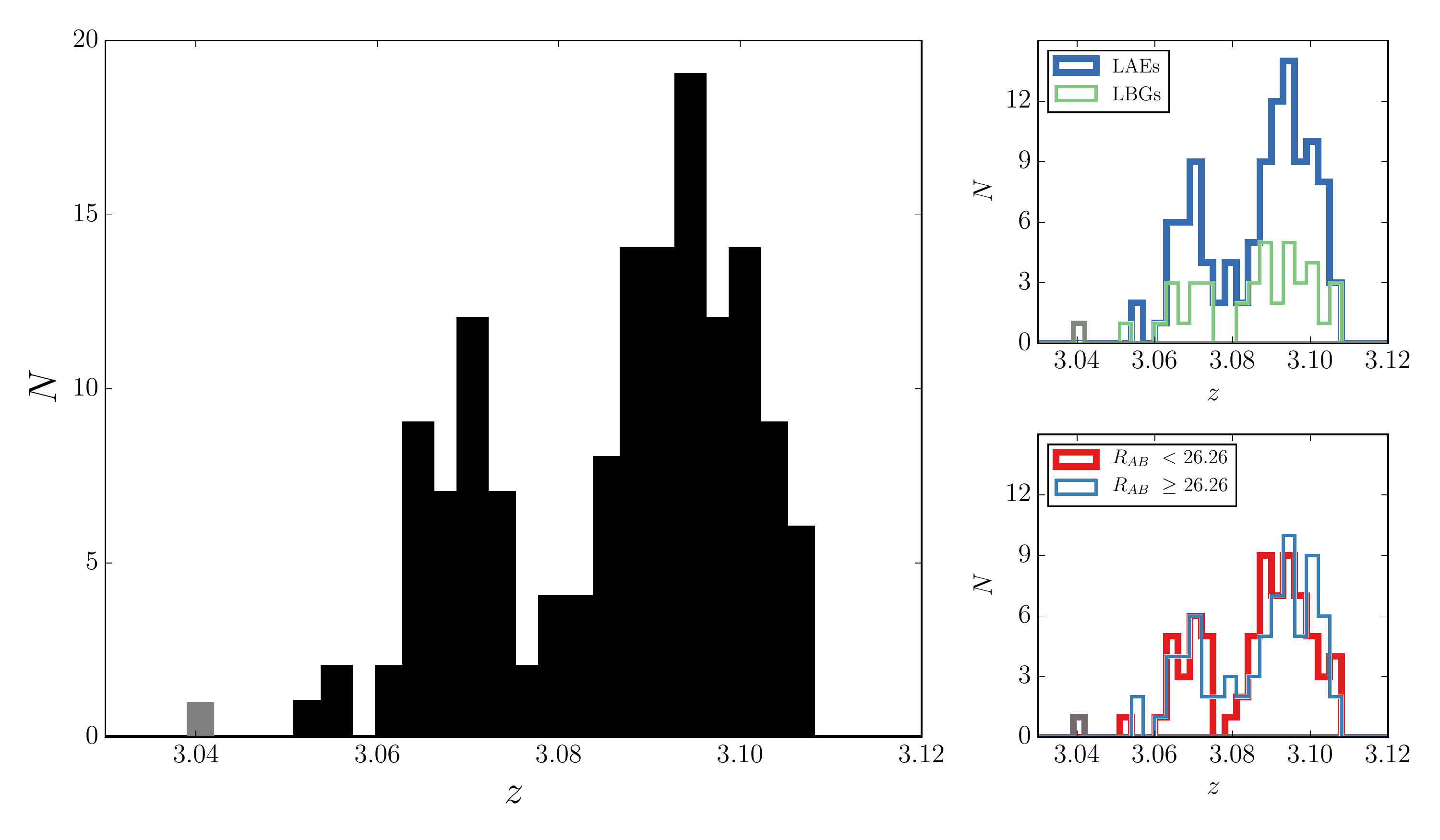}
\caption{{\bf Left:} Redshift histogram of objects within the $z\sim 3.09$ overdensity in the SSA22
field.  This histogram exhibits a double-peaked structure, with one peak centered at $z=3.069$
and the other at $z=3.095$.  The histogram contains a total of 146 galaxies comprising 106 LAEs and 40 LBGs,
plus one additional LBG at $z=3.04$ to show the larger-scale environment. {\bf Top right:}
Distributions of LAEs (blue histogram) and LBGs (green histogram),
both of which appear to trace two distinct concentrations in redshift space. {\bf Bottom right:}
Distributions of brighter $(R_{AB} < 26.26)$ (red histogram) and fainter $(R_{AB}\ge 26.26)$ (blue
histogram) galaxies.  We find no significant difference in the redshift distributions of the brighter and fainter
galaxies.}
\label{fig:histogram-all}
\end{figure*}

\section{Observations \& Methods}
\label{sec:obsAndMethods}
\subsection{Data}
\label{sec:data}
In the analysis presented here, we consider the redshifts and sky
positions of both LAEs and LBGs in the SSA22 field.
The LAEs were selected to lie at $3.05 \leq z \leq 3.12$ using deep broadband BV imaging from 
Subaru/Suprime-cam and narrowband $4980\rm \AA$ imaging from 
Keck/LRIS and Subaru/Suprime-cam.  The LBGs were drawn from the large survey of
\citet{Steidel2003}, in which LBGs were identified using deep $U_nG{\cal R}$ imaging and 
photometric selection criteria tuned to find star-forming galaxies at $z\sim3$.  
Our key results are based on dense spectroscopic sampling of galaxies over a $5.5' \times 7.6'$
region within the original ``SSA22a" pointing of \citet{Steidel1998}, centered on R.A. = 22:17:34,
decl.=00:15:04 (J2000).  Keck/LRIS spectroscopy for the LBGs in our sample was obtained over the course of
several observing campaigns utilizing different instrumental setups.  
Within the region of dense spectroscopic sampling, redshifts have been measured for 85\% of LBG photometric
candidates and 82\% of LAE narrowband-selected candidates.
A full description of these imaging and spectroscopic
observations, including the methods of data reduction, can be found in
\cite{Steidel2003} and \citet{Nestor2011, Nestor2013}.

\subsection{Redshift Measurements}
\label{sec:zMeasurement}

We measured redshifts for objects in our sample based on the observed
wavelengths of Ly$\alpha$ emission, and, if present, interstellar metal absorption lines. 
For measuring redshifts, we developed code to systematically fit profiles 
to either emission or absorption lines.  
Ly$\alpha$ emission profiles were initially fit with single Gaussian functions.
In the case of Ly$\alpha$ emission exhibiting a double-peaked morphology, 
however, a single Gaussian did not provide an adequate description of the profile,
and an additional Gaussian was required to obtain an acceptable fit.
Galaxies with double-peaked Ly$\alpha$ emission in our sample typically show one of two different
morphologies, with either two peaks of comparable amplitude or else a significantly stronger red peak
and weaker blue peak.  We used the same method to measure the redshift for both types of
double-peaked Ly$\alpha$ emission, based on the wavelength of the center of the trough between
the two peaks. To estimate the interstellar absorption redshift in LBG spectra, 
we averaged the redshift measurements from individual absorption lines.
In our sample, some objects had repeat observations. In such cases, we adopted
the redshift measured from the observation with the highest signal to noise ratio (SNR).  

In total, we measured redshifts for 202 galaxies, including 116 LAEs and 86 LBGs.  
Due to our interest in the $z\sim3.09$ protocluster, we now focus only on galaxies 
at $3.05 \le z \le 3.12$. 
This redshift cut reduces our sample to 146 galaxies, 
comprising 106 LAEs and 40 LBGs.

Ly$\alpha$ emission and interstellar absorption features typically trace gas that is outflowing
from galaxies, which perturbs their measured redshifts from the systemic value.
In order to investigate the large-scale structure traced by LBGs and LAEs in SSA22,
we need to estimate the systemic redshift of each galaxy.  We obtain
systemic redshifts for objects in our sample by shifting the measured Ly$\alpha$
and interstellar absorption redshifts to the rest frame of the galaxies. 

The shift required to estimate the systemic redshift has been measured to be 
different for LAEs and LBGs, and depends on which spectral features are 
observed.  Based on the results of \citet{Trainor2015}, we shifted the redshifts of objects 
classified as LAEs with only Ly$\alpha$ emission by $\delta v=-200 \rm \ km \ s^{-1}$, and the
redshifts of LAEs showing both Ly$\alpha$ emission and interstellar absorption 
lines by $\delta v = (0.114\times \Delta v_{abs,em} - 230 \rm \ km \ s^{-1})$, 
where $\Delta v_{abs,em}$ is the velocity difference between the Ly$\alpha$ 
and interstellar absorption redshifts.  Based on equations presented in \citet{Adelberger2003},
we shifted the redshifts of objects classified as LBGs showing Ly$\alpha$ 
emission by $\delta v = -310 \rm \ km \ s^{-1}$, 
and finally, the redshifts of LBGs showing only interstellar absorption 
lines by $\delta v = 150 \rm \ km \ s^{-1}$.  To apply 
the systemic redshift correction, we converted the shifts from velocity to
redshift space using $\delta v/c = \delta z/(1+z)$.  Following these 
corrections, we obtained a final list of LAE and LBG systemic redshifts.

\section{Results}
\label{sec:results}
\subsection{The SSA22 Redshift Distribution}
\label{sec:histogram}

We constructed a histogram from the finalized list of 146 systemic redshifts in the 
SSA22 field (Figure~\ref{fig:histogram-all}).  This histogram shows that
the spike discovered in \citet{Steidel1998} at $z\approx3.09$ contains 
two distinct peaks, one with a central redshift of $z_{peak,b}=3.069$ (blue peak), and 
another centered on $z_{peak,r}=3.095$ (red peak).    
The peaks are separated by $\Delta v\sim 1900 \mbox{ km s}^{-1}$ and $\sim 24$ comoving
Mpc along the line of sight.  We fit a Gaussian to each of the two peaks, defining the boundary between the two 
peaks as the bottom of the trough of the distribution ($z=3.0788$).  Accordingly,
the blue and red peaks contain, respectively, 43 and 103 galaxies.
In velocity space, the standard deviations of the blue and red peaks are, respectively: 
$\sigma_{v,b}= 350 \rm \ km \ s^{-1}$ and $\sigma_{v,r} = 540\rm \ km \ s^{-1}$. Taking
into account the uncertainties associated with estimating systemic redshifts 
\citep[$\sim 170\mbox{ km s}^{-1}$;][]{Adelberger2003,Trainor2015},
we find the intrinsic widths to be 310 and 520 $\rm \ km \ s^{-1}$, respectively, for the blue and red peaks. 
We note that the velocity width and centroid of the red peak are similar to values found by
\citet{kubo2015} for a sample of $K$-selected galaxies in the SSA22 overdensity with
rest-frame optical spectroscopic measurements.
The properties of the two peaks are summarized in Table~\ref{table:hist-parameters}.

\begin{deluxetable*}{ccccc}
\tabletypesize{\footnotesize}
\tablecolumns{5}
\tablewidth{0pt}
\tablecaption{Redshift Histogram Fit Parameters}
\tablehead{
\colhead{Peak} & \colhead{$N_{gal}$} & \colhead{$z_{peak}$} & \colhead{$\sigma_z$\tablenotemark{a}} & \colhead{$\sigma_v$ (km s$^{-1}$)\tablenotemark{a}} }
\startdata
Blue &  43 & 3.069 $\pm$ 0.001 & (4.74 [4.25] $\pm$ 0.72) $\times \ 10^{-3}$ & 350 [310] $\pm$ 53 \\
Red & 103 &3.095 $\pm$ 0.001 & (7.37 [7.12] $\pm$ 0.54) $\times \ 10^{-3}$ & 540 [520] $\pm$ 40
\enddata
\tablenotetext{a}{Values in brackets result after deconvolving the uncertainties in systemic redshifts
\citep{Adelberger2003,Trainor2015} from the observed sample standard deviations in redshift and velocity space.}

\label{table:hist-parameters}
\end{deluxetable*}

We investigated the relative distributions of LAEs and LBGs in the two peaks.  
As shown in Figure~~\ref{fig:histogram-all}, both the
LAE and LBG distributions show evidence of double peaked structure.  
LBGs [LAEs] comprise 28\% [72\%] of the blue peak, and 27\% [73\%] of the red peak.   
A Kolmogorov-Smirnov (K-S) test indicates a probability of 98\% that the LAE
and LBG redshift distributions are drawn from the same parent distribution.

We also considered how brighter and fainter galaxies are distributed within the SSA22 stucture.
All LAEs and LBGs are covered by a deep Subaru/Suprime-Cam R-band image \citep{Hu2004}, ranging in magnitude
from $R_{AB}=21.61$ (a QSO) to limits fainter than $R_{AB}=27.0$ (roughly half of the LAEs). 
We classified each galaxy as either brighter or fainter, 
depending on its magnitude relative to the 
sample median of $R_{AB}=26.26$. As shown in Figure~\ref{fig:histogram-all}, the brighter
and fainter histograms are very similar. A K-S test indicates a probability
of 87\% that the brighter and fainter redshift distributions are drawn from the same parent
distribution.  Galaxy correlation functions of $z\sim 3$ LBGs suggest that more UV-luminous galaxies are more 
strongly clustered \citep{Ouchi2004, Adelberger2005}. In future work, we will compare the
relative distributions of brighter and fainter galaxies (i.e., more and less massive dark matter halos)
in cosmological simulations of protoclusters \citep[e.g.,][]{Klypin2011} with those observed in the SSA22 overdensity.

\citet{Lehmer2009b} and \citet{Alexander2016} identify
eight X-ray AGNs in the SSA22 field with spectroscopic
redshifts falling within the protocluster range. We used the X-ray catalog of \citet{Lehmer2009a} to
find two additional matches with spectroscopically-confirmed LAEs in our catalogs 
\citep[LAE017 at $z=3.105$ and LAE076 at $z=3.066$;][]{Nestor2013},
for a total sample of 10 X-ray AGN at $z=3.06-3.11$. Nine out of these 10 AGNs fall within the red peak of the SSA22 redshift
histogram (and five out of 6 within the field of view of our dense spectroscopic sampling).
Although there are more LBGs and LAEs identified in the higher-redshift peak (103, vs. 43 in the lower-redshift peak),
the AGN distribution is even more skewed towards higher redshift. \citet{Lehmer2009b} previously
noted an enhanced AGN fraction in the SSA22 protocluster, but now we can specifically associate this enhancement
with the red peak at $z=3.095$. We also note that all seven giant LABs from \citet{Matsuda2011} with
spectroscopic redshifts are located in the red peak.

\begin{figure}[t!]
\centering
\includegraphics[width=0.5\textwidth]{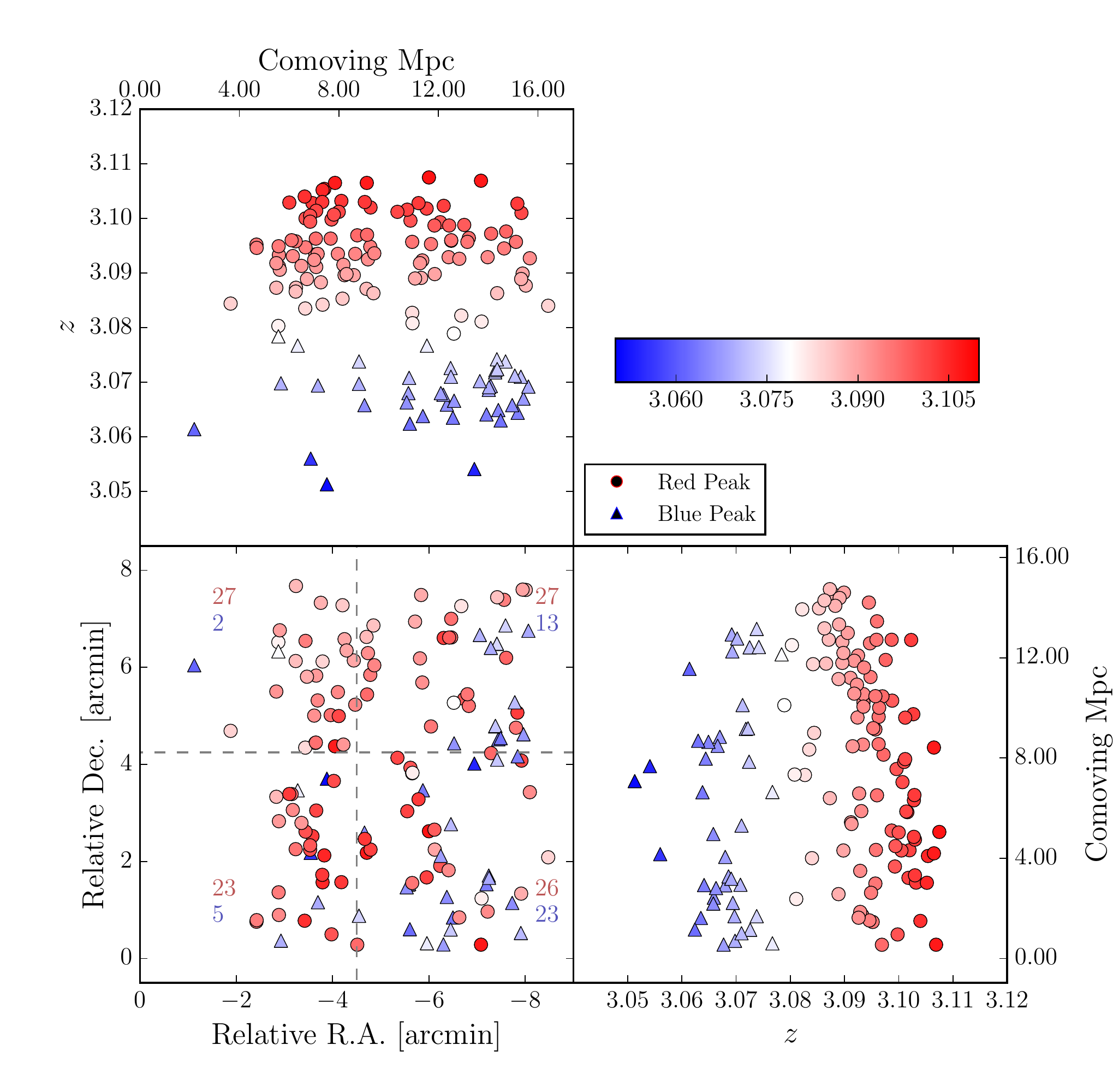}
\caption{Sky and redshift positions of our spectroscopic sample.  Sky coordinates are listed relative to the
southeast corner of the field. Blue triangles [red circles] represent galaxies contained in the redshift peak
centered at $z=3.069$ [$z=3.095$]. The shade of each point corresponds
to redshift of the galaxy being plotted (see colorbar). White on the colorbar corresponds to the bottom of the trough between the two
peaks in the redshift histogram. {\bf Bottom left:}  Galaxies plotted as decl. vs. R.A.
The field is separated into four quadrants by the grey dashed lines.
The numbers of galaxies in each quadrant in the blue and red peaks are listed, respectively, with blue and red numbers.
The region of dense spectroscopic sampling extends over $5.5' \times 7.6'$.
{\bf Top:} Galaxies plotted as $z$ vs. R.A..  Here the absence of galaxies between the blue and red peaks is 
clearly visible, as is the segregation of objects in the blue peak towards the western half of the 
field. {\bf Bottom right:}
Galaxies plotted as decl. vs. $z$.  There is no significant difference in the distributions of blue and red galaxies in decl.}
\label{fig:scatter}
\end{figure}

\subsection{Sky Positions}
\label{sec:scatter}

We use the sky positions of SSA22 LAEs and LBGs to investigate
the spatial distribution of galaxies within the observed redshift
structure. At $z=3.09$, for the assumed cosmology, one arcsecond corresponds
to $7.81$~kpc (proper). Our observations span $\approx 15$~comoving Mpc 
on a side, covering the densest region but not the full extent of the structure
mapped out by \citet{Matsuda2005} and \citet{Yamada2012a} (see \S\ref{sec:discussion}).
The redshift interval $z=3.05-3.12$ corresponds to a Hubble-flow distance of $\sim 66$ comoving 
Mpc, but the translation between redshift and distance is affected by peculiar velocities.

Quite strikingly, we find differences in the spatial distributions of galaxies located
in the two redshift peaks.  Figure~\ref{fig:scatter} displays the sky positions of objects
in our spectroscopic sample, indicating members of blue and red redshift peaks
with, respectively, blue and red symbols. In order to search for a spatial separation of 
galaxies in different redshift peaks, we split the positions of objects in our sample into 
quadrants, represented by grey dashed lines in Figure~\ref{fig:scatter} (bottom left), and
count the number of galaxies in each quadrant that are within either blue or red redshift peaks.
Most significantly, we find that galaxies in the blue peak concentrate towards the western
half of the field, with the northeast quadrant showing the strongest deficiency in
lower-redshift galaxies. This result suggests that galaxies in blue and red redshift peaks 
do not span the same physical volume, and that we are observing the edge of the 
structure containing the galaxies in the blue peak.

\section{Discussion}
\label{sec:discussion}
We have measured redshifts for $146$ LAEs and LBGs at $z=3.05-3.12$ in the SSA22 field,
which clearly delineate two separate peaks in redshift space and also show segregation on the sky.
Collectively, these observations suggest that the the two redshift peaks correspond
to physically distinct structures. 

\begin{figure*}[ht]
\centering
\includegraphics[width=0.95\textwidth]{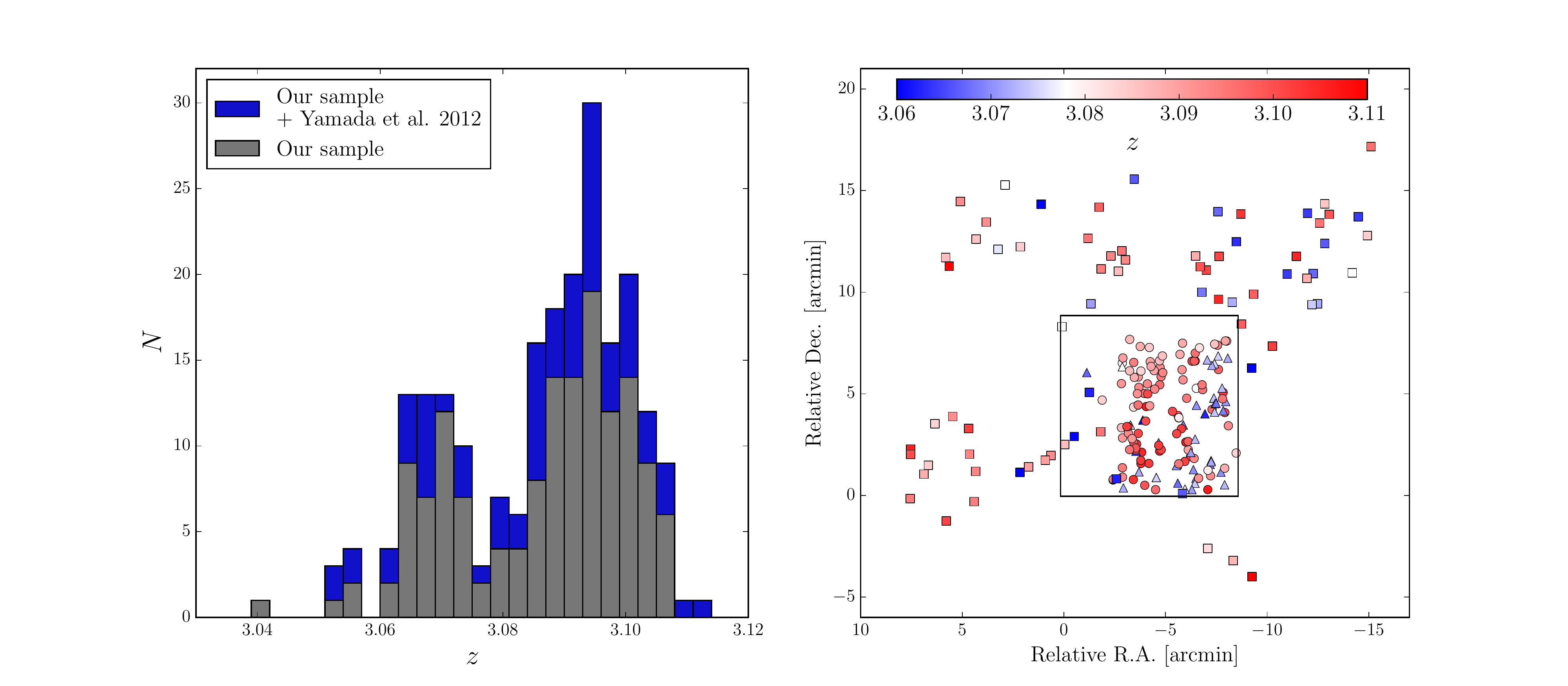}
\caption{{\bf Left:} Redshift histogram of our sample, combined with the sample of \citet{Yamada2012b}.  The double-peaked structure 
is present in the combined distribution, as well as in the sample of \citet{Yamada2012b} alone.
{\bf Right:} Sky positions of the combined SSA22 sample.
Symbols are as in Figure~\ref{fig:scatter} for galaxies in our sample, with
square points for galaxies from  \citet{Yamada2012b}, and color-coding as in Figure~\ref{fig:scatter}.
The original SSA22a pointing from \citet{Steidel1998} is indicated in black.
The relative lack of lower-redshift galaxies (blue points) toward the east side of the 
field is now established over a wider area.} 
\label{fig:yamData}
\end{figure*}

As stated previously, our spectroscopic observations
do not cover the full area of the overdensity in LAEs traced by \citet{Matsuda2005}
and \citet{Yamada2012a}. Therefore, with our data alone, we cannot establish the spatial extent of the double-peaked
redshift histogram. However, we can start to address this question with the observations of \citet{Yamada2012b},
who survey $4980 \rm \ arcmin^2$ in the SSA22 field with Subaru/FOCAS spectroscopy and find 91 
LAEs at a redshift consistent with the $z\sim3.09$ structure.
There are 19 LAEs in common with our sample, resulting in 72 additional objects over a significantly
larger area on the sky (though with sparser sampling). As shown in Figure~~\ref{fig:yamData},
in both the \citeauthor{Yamada2012b} sample alone, and the total sample combined with our data, we detect a double-peaked structure
in redshift. This result suggests that the double-peaked structure is not localized
to the field of view of our observations. We also observe a relative lack of blue-peak galaxies towards
the eastern portion of the field in the combined sample, consistent with our results over a smaller field.

The large-scale structure of the SSA22 overdensity has been considered by \citet{Matsuda2005}, who identify
three filaments of length $\sim 30$ comoving Mpc intersecting within the highest-density region of LAEs on the sky,
and at $z=3.094$ in redshift space. One of these apparent filaments is traced by a concentration of galaxies at $z=3.074$,
which corresponds to the blue peak of our measured histogram.\footnote{Given that the majority of objects
in our histogram are LAEs, whose redshifts were shifted by $\Delta z\sim -0.003$ to the systemic frame, 
our measured $z_{peak,b}=3.069$ would appear at a redshift of $3.072$ in the absence of Ly$\alpha$
velocity corrections, which were not applied by \citeauthor{Matsuda2005}).} With much sparser spectroscopic
sampling, \citeauthor{Matsuda2005} describe this lower-redshift concentration as connecting smoothly
with the large belt-like structure of LAEs at $z=3.088-3.108$ that extends $\sim 40$ comoving 
Mpc across the sky in the northwest -- southeast direction, and corresponds to the red peak of our redshift histogram.
In contrast, our well-sampled redshift distribution indicates that the two peaks are distinct structures that do not join
smoothly in redshift space or fully overlap on the sky, leading to a qualitatively different description
of the large-scale structure in the SSA22 overdensity at $z<3.08$.

\begin{figure}[ht]
\centering
\includegraphics[width=0.5\textwidth]{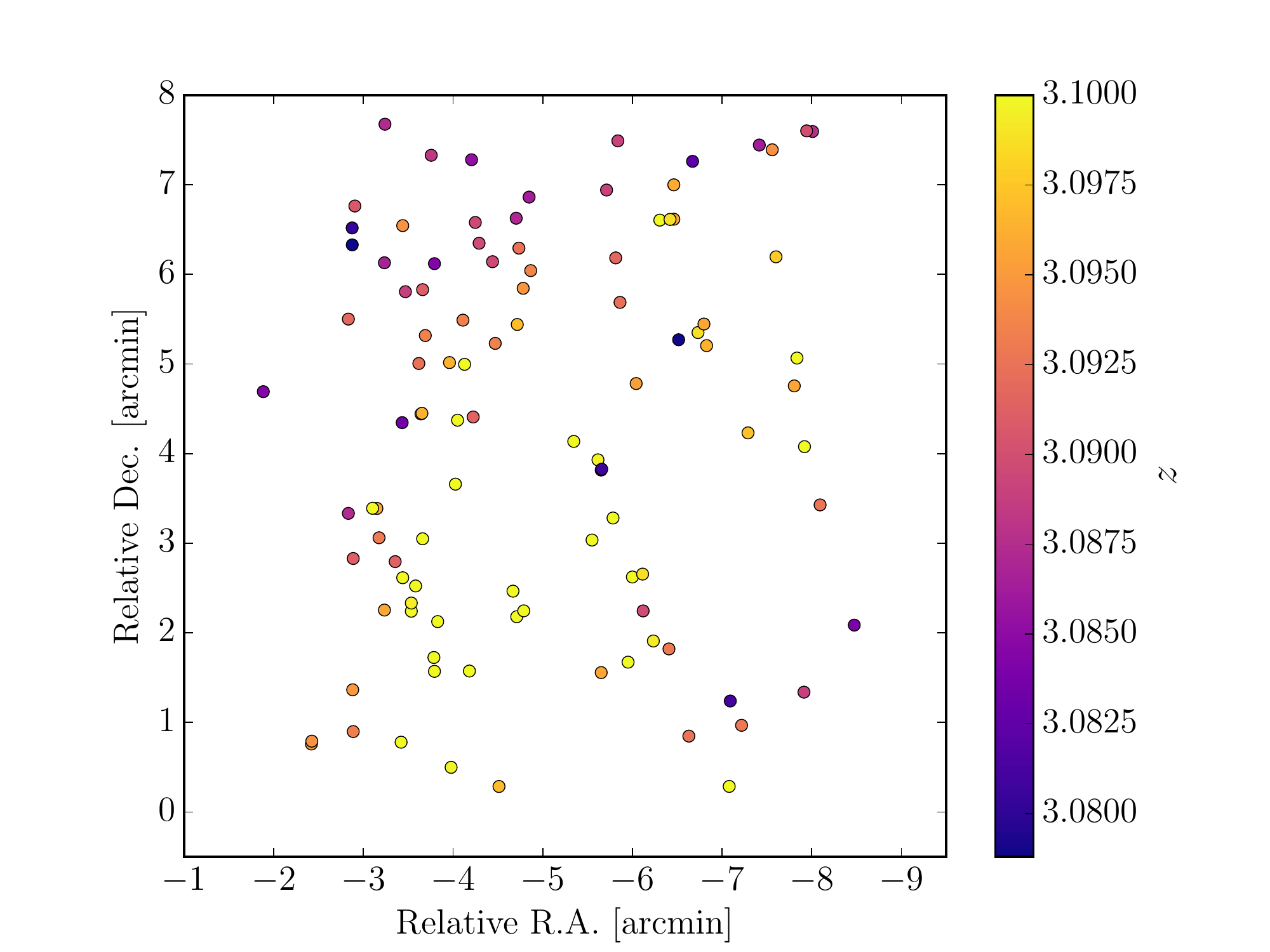}
\caption{Sky positions of objects in the red peak of the SSA22 redshift histogram ($z=3.08-3.11$). Points are color coded
by redshift as indicated in the color bar, and follow a gradient in redshift space from the southwest (lower right, higher redshift)
to northeast (upper left, lower redshift).}
\label{fig:scatterR}
\end{figure}

\citet{Matsuda2005} also present a gradient in redshift from northwest (higher redshift) to southeast (lower redshift)
within the extended band of LAEs at $z=3.088-3.108$. As shown in Figure~\ref{fig:scatterR},
within our smaller field of view, we find a redshift gradient in the red peak ($z=3.08-3.11$) from
southwest (higher redshift) to northeast (lower redshift), i.e., rotated by 90 degrees with respect to the gradient
presented in \citet{Matsuda2005}. Our dense spectroscopic sampling covers only $\sim 30$\% of the linear dimension
of the band of LAEs, so we cannot establish the extent of the redshift gradient we observe. However, based on higher
spatial resolution, our sampling of redshifts as a function of position within the red peak reveals a more complex
picture than the linear gradient presented by \citet{Matsuda2005}.

In order to understand the full three-dimensional architecture, dynamics, origin, and fate of the substructure
we have uncovered with our spectroscopic survey of the SSA22 overdensity, we must obtain comparably
dense spectroscopy over a significantly larger area -- at the very least within the $27' \times 34'$ Subaru/Suprime-Cam
pointing \citep[SSA22-Sb1 from][]{Yamada2012a} containing the initial SSA22a field of \citet{Steidel1998}. 
The spectroscopic samples of \cite{Matsuda2005} and \citet{Yamada2012b}
begin to address this need, but are both incomplete in terms of areal coverage of the field,
and too sparsely sampled within the areas covered. Detailed comparisons with cosmological simulations of large-scale
structure formation \citep[e.g.,][]{Springel2005,Klypin2011} are also required to understand whether
the double-peaked substructure in SSA22 will coalesce by $z=0$ or remain as two distinct clusters.
We will perform such comparisons in future work.

\section*{Acknowledgements}
We thank Abhimat Gautam and Tommaso Treu for useful comments. 
CCS acknowledges support from NSF grants AST-0908805 and AST-1313472.
AES acknowledges support from the David \& Lucile Packard Foundation.
We wish to extend special thanks to those of Hawaiian ancestry on
whose sacred mountain we are privileged to be guests. Without their generous hospitality, most
of the observations presented herein would not have been possible.


\clearpage

\end{document}